\documentclass[runningheads,a4paper]{llncs}

\usepackage[T1]{fontenc}
\usepackage{amsmath,amssymb}
\usepackage{graphicx}
\usepackage{booktabs}
\usepackage{array}
\usepackage{placeins}
\usepackage{url}
\usepackage[hidelinks]{hyperref}
\usepackage{xcolor}
\usepackage{tikz}
\usetikzlibrary{arrows.meta,positioning,fit,backgrounds,calc,shapes.geometric}
\usepackage{pgfplots}
\pgfplotsset{compat=1.18}

\usepackage{needspace}
\usepackage{listings}
\newcommand{\tr}{\ensuremath{t_{r}}}          
\newcommand{\td}{\ensuremath{t_{d}}}          
\newcommand{\ta}{\ensuremath{t_{a}}}          
\newcommand{\Ae}{\ensuremath{A_{e}}}          
\newcommand{\Fmax}{\ensuremath{F_{\max}}}     
\newcommand{\MR}{\ensuremath{M_{R}}}          
\newcommand{\MF}{\ensuremath{M_{F}}}          
\newcommand{\Dbud}{\ensuremath{D}}            
\newcommand{\Eideal}{\ensuremath{\mathcal{A}_{\mathrm{ideal}}}}
\newcommand{\Ehil}{\ensuremath{\mathcal{A}_{\mathrm{HI}}}}
\newcommand{\phase}{\ensuremath{\phi}}        

\makeatletter
\renewcommand\paragraph{\@startsection{paragraph}{4}{\z@}%
  {8\p@\@plus2\p@\@minus1\p@}{-0.6em}{\normalfont\normalsize\itshape}}
\makeatother

\begin{document}

\title{Mission-Aware Attestation Envelopes for\\
       Time-Critical Autonomous Action:\\
       A Hardware-in-the-Loop V2I Study\thanks{This is a preprint of an article accepted for publication in Modelling and Simulation for Autonomous Systems (MESAS 2026), Springer LNCS.}}

\author{Dimitrios Nikou\inst{1} \and Nikolaos Kekatos\inst{2} \and
        Sophia Petridou\inst{3} \and Stylianos Basagiannis\inst{1}}
\titlerunning{Mission-Aware Attestation Envelopes}
\authorrunning{D. Nikou et al.}
\institute{International Hellenic University, Greece \and
           Clone Systems, Cyprus \and
           University of Macedonia, Greece}

\maketitle

\begin{abstract}
An autonomous system that asks for a privileged physical action is usually
gated on integrity evidence: a platform proves what it is running, and the
request is granted or refused on that basis. Such a gate is normally treated as a
predicate, yet the evidence behind it has an age, the decision that consumes it
has a latency, and the physical system that waits for it has a deadline. In this
paper, we formulate mission-aware attestation as a runtime assurance contract
that holds only when integrity is valid, the evidence is fresh enough, and the
decision completes inside a budget derived from the current physical state. The
contract yields four operational outcomes where a binary gate yields two, and it
separates a refusal caused by tampering from one caused by evidence that had
already aged and from one caused by a decision that arrived too late. We evaluate
it on a hardware-in-the-loop vehicle-to-infrastructure platform in which a
driving simulator supplies the physical state and the authorisation deadline,
while a microcontroller on-board unit and a TPM-backed roadside unit running
Linux integrity measurement supply the assurance evidence. Combining
simulator-derived mission state with measured assurance behaviour lets us
construct the attestation envelope in which a security verdict is still
actionable, and compare the region a security-blind model admits against the
region a hardware-informed one admits. The first admits the whole space and the
second three quarters of it, and every point it refuses fails the freshness
margin rather than the response margin. Moving the attestation
interval across the range the verifier itself permits costs about as much as a
fivefold scaling of the latency distribution, and the interval is directly configurable
rather than a property of the hardware, which makes it the immediately actionable
deployment parameter. We also show that the two margins interact: if the freshness bound
does not exceed the authorisation budget, every late decision is also stale and
lateness becomes unobservable, which makes the attestation interval and the
freshness bound design parameters that cannot be chosen from security
requirements alone.

\end{abstract}

\section{Introduction}
\label{sec:intro}

Intelligent transportation systems now place a substantial amount of
decision-making in vehicles and in the roadside infrastructure they talk to, and
vehicle-to-infrastructure (V2I) exchange is the channel through which that
coordination happens~\cite{coppola2016}. Such exchanges let a vehicle request a change to infrastructure it does not own,
of which emergency-vehicle signal preemption is the canonical
example~\cite{fhwa2008,evpdelay2018}, and the security services protecting those
messages are standardised~\cite{ieee1609}. Automating them enlarges an attack surface
already well catalogued for automated vehicles~\cite{petit2015}, and
communication delay is an established input to preemption
performance~\cite{evpdelay2018}. The privileged action, in other words, is granted
to a remote party over a shared channel, on evidence that party supplies.

The mechanism that makes such a grant defensible is remote attestation: a
platform produces evidence about what it is running, and a verifier decides
whether to believe it~\cite{rfc9334}. On general-purpose hosts that evidence is
rooted in a Trusted Platform Module (TPM)~\cite{tcg2019,parno2010} and produced
by the Linux Integrity Measurement Architecture (IMA)~\cite{ima2004}, with
continuous verification against a policy provided by frameworks such as
Keylime~\cite{keylime2016,keylimedsn2025}. On constrained embedded devices, where a discrete
root of trust is often absent, software and hybrid alternatives range from
scalable collective attestation~\cite{seda2015} to control-flow attestation of
the running program~\cite{cflat2016}. In each case a verdict is produced, and
systems built on top consume it as a predicate.

Treating it as a predicate is where the difficulty begins, and the attestation
literature has already identified the pieces without assembling them. Freshness
has been foundational to remote attestation since the area was set out
systematically~\cite{coker2011}, and the architecture standardised for it treats
the freshness of evidence and of attestation results as a matter of policy,
discussing expiry explicitly~\cite{rfc9334}, so age is a recognised dimension. Work on real-time
mission execution integrity argues that autonomous cyber-physical systems need
attestation that is timely as well as correct~\cite{ari2023}, so timeliness is
recognised too. The conflict between security mechanisms and timing guarantees
is a mature line in real-time systems~\cite{contego2017}. What is missing is the
coupling: none of this asks whether a verdict is still \emph{decision-valid} for
a physical action whose admissible response time is set, at run time, by where
the requester is and how fast it is moving.

Three distinct requirements therefore have to be satisfied before a privileged
action is authorised. The evidence must be \emph{valid}, meaning integrity measurement
reports no violation; \emph{fresh}, meaning it describes the platform recently
enough to be informative about it now; and the decision must be \emph{timely},
landing while the physical system can still act on it. A cryptographically valid
quote can be stale; a fresh quote can arrive too late; a timely result can report
a violation. Only the conjunction permits the action, and a binary gate collapses
three distinct reasons for refusal into one, or reports trusted on evidence aged
past usefulness.

\paragraph{Approach.}
We treat the gate as a runtime assurance contract with two margins. The physical
state at a request supplies an authorisation budget through a mission-specific
deadline function, giving a response margin, and the timestamp of the attested
evidence gives a freshness margin against a configured bound. An authorisation is
assurance-valid only when integrity passes and both margins are non-negative, so
the contract has four outcomes rather than two and the outcome names the reason.
Runtime verification supplies the framing for monitoring such obligations over an
execution~\cite{bauer2011,falcone2021} and its extension into
cyber-physical settings is an active concern~\cite{sanchez2019}; what we add is
that one bound of the obligation is fixed by physical state at the moment the
obligation is created.

\paragraph{The role of simulation.}
The deadline is a function of physical state, so something has to supply that
state across the operating space. A driving simulator does, at speeds and
geometries impractical to instrument on a road~\cite{carla2017}. The assurance behaviour cannot be obtained the same way,
because its distribution depends on a real TPM, a real integrity-measurement
stack and a real network. Combining the two in a hardware-in-the-loop (HIL)
arrangement, the established way of exercising real components against a
simulated plant~\cite{isermann1999}, is what lets us ask where in the simulated
operating space the real security mechanism can meet its contract, and the answer
is a region rather than a number. We describe the simulator as a simulator: it is not synchronised with
or calibrated against a physical intersection, so it is not a digital
twin~\cite{nistdt2021}.

\paragraph{Contributions.}
\begin{enumerate}
\item We formulate mission-aware attestation as a runtime assurance contract
requiring integrity validity, evidence freshness and completion before a
state-dependent physical deadline, with four operational outcomes in place of a
binary verdict, and we show the two margins interact: below a threshold on the
freshness bound, lateness is unobservable (Sect.~\ref{sec:contract}).
\item We give a simulation and HIL methodology that combines simulator-derived
mission state with measured TPM, IMA and verifier behaviour to construct an
attestation envelope, and that compares the envelope admitted by a
security-blind model against that admitted by a hardware-informed one
(Sects.~\ref{sec:platform} and~\ref{sec:res:envelope}).
\item We characterise the envelope empirically and show that the attestation
interval is a design parameter of comparable weight to decision latency, and one
that a deployment can change without touching the hardware
(Sect.~\ref{sec:results}).
\end{enumerate}

\paragraph{Research questions.}
The evaluation asks four.
\begin{description}
\item[RQ1] How much of the authorisation budget does the assurance loop consume,
and does channel contention change it?
\item[RQ2] Over what region of trigger distance, approach speed and evidence age
is the integrity verdict still actionable?
\item[RQ3] How much does a security-blind model of the physical process
overestimate that region?
\item[RQ4] How sensitive is the envelope to the attestation interval, against
latency?
\end{description}

\paragraph{Scope.}
Latency is reported as a measured distribution and never as a worst-case bound,
and the deadline enforced is an authorisation-decision budget within the
experimental scenario rather than the full preemption budget a fielded controller
would need~\cite{fhwa2008}. Sect.~\ref{sec:threats} sets out the remaining
limits.

\section{System Model and Assurance Problem}
\label{sec:scenario}

This section fixes the object the rest of the paper reasons about: the scenario
and the loop realising it, an abstract model of components, events and timing,
the trust and threat assumptions, and the failures a binary gate cannot report.
Implementation follows in Sect.~\ref{sec:platform}.

\subsection{Operational V2I Scenario}
\label{sec:opscenario}

The scenario is a time-critical authorisation: an approaching vehicle requests a
privileged change to infrastructure it does not control, and the infrastructure
must decide before the vehicle arrives. We instantiate it as signal preemption,
an emergency vehicle requesting right of way at an intersection. The instantiation is concrete, but the
structure is generic to privileged-access authorisation: a requester with an
identity, a decision point with its own integrity to account for, an actuation
gate, and a physical process that fixes how long the decision may take. The same four elements
arrange the same way in an unmanned ground vehicle requesting barrier release at
a checkpoint, or any case in which an approaching system asks infrastructure it
does not own to change state on evidence it supplies. What differs across those
settings is the cost of a wrong decision and how far the channel can be
contested, and both raise the value of knowing where evidence is still
actionable.

\begin{figure}[tbp]
\centering
\resizebox{\linewidth}{!}{%
\begin{tikzpicture}[font=\scriptsize,>=Latex,
  lane/.style={draw,fill=black!4,minimum width=2.3cm,minimum height=9mm,
               text width=2.1cm,align=center,inner sep=1pt},
  msg/.style={->,thick},
  note/.style={font=\scriptsize\itshape,text=black!60}]

\foreach \i/\n in {0/{Simulator\\ \textit{physical state}},
                   1/{OBU\\ \textit{identity}},
                   2/{Broker\\ \textit{transport}},
                   3/{RSU\\ \textit{decision point}},
                   4/{TPM $+$ IMA\\ \textit{evidence}}}
  \node[lane] (L\i) at (2.7*\i,0) {\n};
\foreach \i in {0,...,4} \draw[black!25,dashed] (2.7*\i,-0.5) -- (2.7*\i,-5.9);

\newcommand{\step}[4]{\draw[msg] (2.7*#1,#3) -- node[above,font=\scriptsize]{#4} (2.7*#2,#3);}

\step{0}{2}{-1.05}{(1) proximity beacon}
\step{2}{1}{-1.50}{}
\step{1}{2}{-2.00}{(2) MAC $+$ SHA-256 firmware hash}
\step{2}{3}{-2.45}{}
\step{3}{4}{-3.05}{(3) status query}
\draw[msg] (2.7*4,-3.50) -- node[above,font=\scriptsize]{verdict, $t_a$} (2.7*3,-3.50);
\step{3}{2}{-4.25}{(4) \textsf{PERMIT} / \textsf{DENY-\{INTEGRITY,STALE,LATE\}}}
\step{2}{0}{-4.70}{}
\step{0}{1}{-5.45}{(5) actuate}

\draw[<->,black!55] (-1.75,-1.05) -- node[left,note,align=right,xshift=1mm]
     {authorisation\\budget $D(d_r,v_r)$} (-1.75,-4.70);
\node[note,anchor=west,align=left] at (2.7*4+1.35,-3.28)
     {refreshed only\\every $T$};
\end{tikzpicture}}
\caption{The authorisation loop, with steps numbered as in the text. Dashed
verticals are component lifelines and time runs downwards.}
\label{fig:loop}
\end{figure}
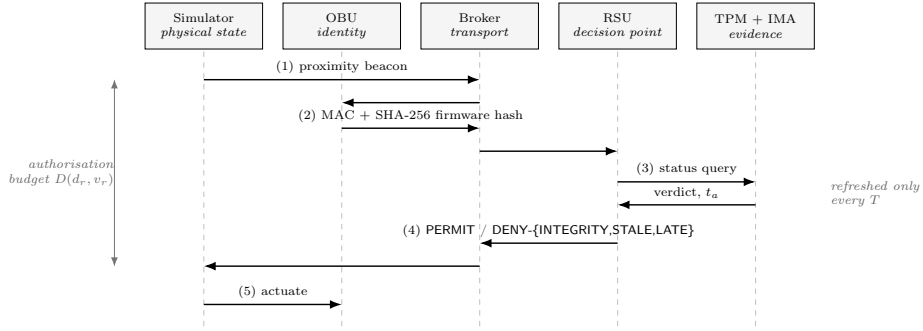

Figure~\ref{fig:loop} gives the loop in five steps. (1)~The simulator detects the
vehicle entering the trigger volume at distance $d_r$ and speed $v_r$, which
together fix the authorisation budget $\Dbud(d_r,v_r)$, the time the decision has
before the vehicle passes the point at which actuation is still usable, and it
emits a proximity beacon. (2)~The on-board unit answers with an admission
payload: its registered hardware address together with a firmware digest
recomputed at that moment. (3)~The roadside unit checks that payload against a
table of admitted units, then queries the verifier that attests its own platform,
because a decision point whose execution has been tampered with can return any
answer at all. That query returns an integrity verdict together with $\ta$, the
time at which the evidence behind it was generated; the evidence is refreshed
only once every attestation interval $T$, so its age at the request may be as
much as $T$, and more by the time the decision is taken. (4)~The contract of
Sect.~\ref{sec:contract} yields one of four outcomes, naming which requirement
failed. (5)~On \textsf{PERMIT} the simulator actuates.

\subsection{System Model}
\label{sec:sysmodel}

The system is the tuple
$\mathcal{S} = \langle \mathcal{C}, \mathcal{E}, \mathcal{Q}, \Theta \rangle$ of
components, events, verification state and timing, defined as follows and used
unchanged for the rest of the paper. The letters $V$ and $T$ are reserved for the
decision point and the attestation interval respectively, so the verification
state is written $\mathcal{Q}$ and the timing model $\Theta$. Table~\ref{tab:notation} collects the
notation.

\paragraph{Components $\mathcal{C}$.}
Four, given here as roles; Sect.~\ref{sec:platform} gives the hardware realising
them. The \emph{requester} $R$ is a constrained embedded
device carried by the vehicle, the on-board unit (OBU). The \emph{decision point}
$V$ is the roadside unit (RSU), a general-purpose host that authorises or refuses
and that has integrity of its own to account for. The \emph{evidence source} $P$
is an attestation subsystem that produces platform measurements about $V$ on a
periodic cycle. The \emph{plant} $W$ is the vehicle and intersection, supplying physical state
and executing the actuation.

\paragraph{Events $\mathcal{E}$.}
A run is the sequence
$\langle \mathit{beacon}, \mathit{req}(r), \mathit{quote}(t_a), \mathit{dec}(r,o), \mathit{act}\rangle$
of Fig.~\ref{fig:loop}. A request carries $r=(\tr, x_r)$, with $x_r=(d_r,v_r)$
the plant state at $\tr$: distance to the actuation point and approach speed. A
decision carries its time $\td$ and an outcome $o$ from the four values of
Sect.~\ref{sec:contract}.

\paragraph{Verification state $\mathcal{Q}$.}
Two independent checks. \emph{Identity} is a per-request predicate over what $R$
presents, evaluated against a table of admitted units, so it is as fresh as the
request. \emph{Platform integrity} is a periodic predicate over $V$ itself,
evaluated by $P$ against a measurement policy; $V$ reads its current value with
$\ta$, when the underlying measurement was generated. Only the second has an
age.

\paragraph{Timing $\Theta$.}
$P$ re-attests on an interval $T$, and $V$ tolerates a maximum interval
$T_{\max} \ge T$ before it treats the evidence source as unresponsive. Both are configuration values, and together
they bound the age of held evidence before decision latency is added. Durations are differences of readings of a single clock at one
observation point, so the model assumes no synchronisation between $R$, $V$ and
$W$.

\begin{table}[tb]
\centering
\caption{System-model notation. The contract of Sect.~\ref{sec:contract} adds
$\Dbud$, $\Fmax$, $\MR$ and $\MF$, properties of the assurance policy rather than
facts about the system.}
\label{tab:notation}
\setlength{\tabcolsep}{10pt}
\renewcommand{\arraystretch}{1.05}
\begin{tabular}{@{}ll@{}}
\toprule
\textbf{Symbol} & \textbf{Meaning} \\
\midrule
$R$, $V$, $P$, $W$ & requester, decision point, evidence source, plant \\
$\tr$, $\td$ & time the request is issued, time the decision is taken \\
$\ta$ & time the integrity evidence used by that decision was generated \\
$x_r = (d_r, v_r)$ & plant state at the request: distance to actuation, speed \\
$T$ & attestation interval \\
$T_{\max}$ & verifier liveness threshold: longest tolerated gap between quotes \\
$\Ae = \tr - \ta$ & evidence age at the request \\
$\phase$ & request phase within the attestation interval, in $[0,1]$ \\
\bottomrule
\end{tabular}
\end{table}

\subsection{Threat and Trust Model}
\label{sec:threat}

With the roles fixed, both what is trusted and what the adversary may reach can
be stated against them. We assume $P$ is sound, that the measurements it produces
about $V$ are authentic, and that $W$ reports the physical state faithfully;
$W$ is trusted because it stands in for physical reality rather than for a
component. Nothing else is assumed trustworthy, and in particular $V$ is not.

The
adversary may operate a rogue OBU and broadcast requests from it; may modify
files on $V$, including those covered by the measurement policy; and may generate
arbitrary traffic on the shared channel, in volume and at will.

The adversary may not forge platform-integrity evidence, find digest collisions,
or compromise $P$ or its policy provisioning. Physical attacks on the evidence
source, supply-chain compromise of the measured software, spoofing or replay of
the requester's credential, and compromise of the plant are outside the model.

The third capability defeats neither identity checking nor platform integrity,
and could nonetheless deny the privileged action by increasing decision latency
and exhausting the response margin $\MR$. The budget $\Dbud(x_r)$ is fixed by
physical state and is not itself attackable. It is the capability whose reach
this paper measures without exercising it.

\subsection{Limitations of a Binary Gate}
\label{sec:blind}

Written as a predicate over a verdict, the gate has two answers, while the scenario
contains four relevant failure cases spanning three distinct causes of refusal.

A \emph{rogue or unenrolled requester} presents an identity absent from the
admitted table, which is what identity checking is for and the gate reports it.
An enrolled requester whose firmware has been replaced is a different matter;
Sect.~\ref{sec:platform} explains why this pillar does not address it. A
\emph{compromised decision point} is subtler, because the deciding platform is
itself a target: an attacker able to modify files on $V$ can change what it
grants.
Binding $V$'s execution to a hardware root of trust is what integrity measurement
adds, and the gate reports that too.

The third failure is a decision that is correct and arrives after the plant has
passed the point at which actuation is usable. The gate returns a verdict and the
verdict is right, and under a fail-secure policy the privileged action is refused
anyway, so the infrastructure stays safe and the requester loses the service.

The fourth is invisible to a binary gate. Platform integrity is periodic, so at
an arbitrary moment the most recent evidence is up to $T$ old, and up to
$T_{\max}$ in the worst case the verifier permits. A gate that reads the current
status reports trusted, and what it means is that $V$ was trusted at some point
within that window. Whether that is adequate is a question about $W$ rather than about the
cryptography.

\section{Mission-Aware Attestation Contract}
\label{sec:contract}

This section defines when integrity evidence is actionable. We separate three
properties that a binary gate conflates, give the contract that requires all
three, and state the outcome semantics that follow from it.

\paragraph{Events and quantities.}

A request event is $r = (\tr, x_r)$, where $\tr$ is the time the requester asks
for the privileged action and $x_r$ the physical state at that moment; in the
instantiation here $x_r = (d_r, v_r)$ is distance to the actuation point and
approach speed. The decision is taken at $\td$. The integrity evidence consumed by
that decision was generated at $\ta$, so its age at the request is
$\Ae = \tr - \ta$.

A mission-specific deadline function maps the physical state to an authorisation
budget, $\Dbud(x_r)$. In the instantiation we take $\Dbud(d_r, v_r) = d_r / v_r$,
which is an upper bound on the budget rather than the budget itself: a fielded
controller must also transmit the command, act on it, and run clearance intervals
before the privileged state is safely reached~\cite{fhwa2008}. We return to this
in Sect.~\ref{sec:discussion}.

\subsection{Authorisation Budget, Evidence Age and Margins}

\begin{figure}[tbp]
\centering
\resizebox{\linewidth}{!}{%
\begin{tikzpicture}[font=\small,>=Latex,
  ev/.style={thick},
  ov/.style={fill=black!12}]

\fill[ov] (7.4,0.55) rectangle (9.2,1.05);
\draw[<->,ev] (0.8,0.8) -- (7.4,0.8);
\node[fill=white,inner sep=1pt] at (4.1,0.8) {$\Fmax$: largest age the policy accepts};
\draw[<->,ev] (0.8,1.55) -- (9.2,1.55);
\node[fill=white,inner sep=1pt] at (5.0,1.55) {evidence age at the decision, $\td-\ta$};
\node[anchor=west] at (9.35,0.8) {overrun $=-\MF$};
\node[anchor=west] at (9.35,0.35) {$\Rightarrow$ \textsf{DENY-STALE}};

\draw[->,thick] (0.2,0) -- (11.0,0) node[right]{time};
\foreach \x/\l in {0.8/{$\ta$}, 6.0/{$\tr$}, 9.2/{$\td$}}{
  \draw[thick] (\x,-0.14) -- (\x,0.14);
  \node[anchor=south,inner sep=2pt] at (\x,0.16) {\l};}
\node[anchor=north,align=center] at (0.8,-0.22) {evidence\\generated};
\node[anchor=north,align=center] at (6.0,-0.22) {request};
\node[anchor=north,align=center] at (9.2,-0.22) {decision};

\fill[ov] (8.5,-1.75) rectangle (9.2,-1.25);
\draw[<->,ev] (6.0,-1.5) -- (8.5,-1.5);
\node[fill=white,inner sep=1pt] at (7.2,-1.5) {$\Dbud(x_r)$};
\draw[<->,ev] (6.0,-2.25) -- (9.2,-2.25);
\node[fill=white,inner sep=1pt] at (7.5,-2.25) {decision latency, $\td-\tr$};
\node[anchor=west] at (9.35,-1.5) {overrun $=-\MR$};
\node[anchor=west] at (9.35,-1.95) {$\Rightarrow$ \textsf{DENY-LATE}};

\draw[black!35,dashed] (9.2,-2.45) -- (9.2,1.75);
\end{tikzpicture}}
\caption{The two ways an authorisation can fail while integrity holds. Above the
axis, the evidence had already aged before the request, and the age it reaches by
the decision overruns the freshness bound $\Fmax$. Below the axis, the evidence
would have been acceptable and the decision latency overruns the authorisation
budget $\Dbud(x_r)$ that the physical state allows. The two are measured from
different origins, $\ta$ and $\tr$, which is why neither implies the other in
general; Sect.~\ref{sec:masking} gives the condition under which one does imply
the other. Shaded spans are the overruns.}
\label{fig:twobudget}
\end{figure}
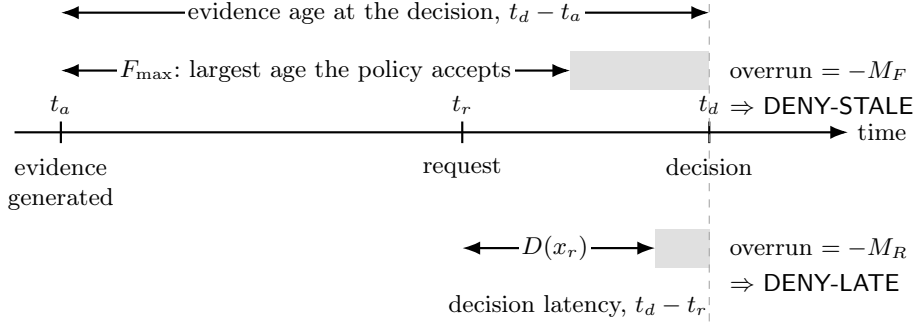

Two margins follow, drawn in Fig.~\ref{fig:twobudget} and measured from
different origins:
\begin{align}
\MR &= \Dbud(x_r) - (\td - \tr) && \text{response margin,}\\
\MF &= \Fmax - (\td - \ta) && \text{freshness margin,}
\end{align}
where $\Fmax$ is the largest evidence age the mission-aware
authorisation policy will accept, a distinct quantity from the verifier liveness
threshold $T_{\max}$ of Sect.~\ref{sec:sysmodel}. Freshness here is bounded
evidence age rather than nonce-established recency: these are the timestamp and
challenge-response variants the attestation architecture
distinguishes~\cite{rfc9334}, and the platform evaluated in
Sect.~\ref{sec:platform} produces evidence on a periodic pull with no nonce
binding a quote to a request. The distinction matters again in
Sect.~\ref{sec:masking}, where the premise of the masking result fails precisely
for evidence obtained by challenge. A negative
$\MR$ means the verdict was correct and arrived too late to be used. A negative
$\MF$ means the system knows only that the platform \emph{was} trustworthy.

\subsection{Contract and Outcome Semantics}
\label{sec:outcomes}

This subsection gives the condition under which an authorisation may be granted,
and the four outcomes that follow from which of its three conjuncts fails.

For the experimental instantiation we adopt the conservative mission-coupled
policy $\Fmax = \Dbud(x_r)$: an authorisation does not rest on an integrity
assertion older than the entire remaining decision horizon. This is a policy
choice rather than a property of remote attestation, and fixed or independently
derived freshness thresholds are equally admissible and would change how the two
margins interact. It has one consequence worth stating: since evidence age at the decision is at least the decision
latency, the freshness test then subsumes the response test, which is why
Sect.~\ref{sec:res:interval} reports what the second margin does and does not do
in this study. A fixed policy constant is the natural alternative and would make
the two margins bind independently.

\begin{definition}[Attestation envelope]
The \emph{attestation envelope} is the subset of the physical and assurance
operating space in which integrity evidence satisfies the contract below and is
therefore actionable.
\end{definition}

An authorisation is \emph{assurance-valid} exactly when
\begin{equation}
\label{eq:contract}
\mathit{integrity}(\ta) = \textsf{PASS}
\;\wedge\;
\MR \ge 0
\;\wedge\;
\MF \ge 0 .
\end{equation}
Equation~\eqref{eq:contract} says an authorisation is refused, not which conjunct
to blame when more than one fails, so outcomes carry a diagnostic precedence:
integrity, then freshness, then timeliness.
\begin{equation}
\label{eq:outcome}
o \;=\;
\begin{cases}
\textsf{DENY-INTEGRITY}, & \mathit{integrity}(\ta) \neq \textsf{PASS},\\
\textsf{DENY-STALE},     & \mathit{integrity}(\ta) = \textsf{PASS} \wedge \MF < 0,\\
\textsf{DENY-LATE},      & \mathit{integrity}(\ta) = \textsf{PASS} \wedge \MF \ge 0 \wedge \MR < 0,\\
\textsf{PERMIT},         & \text{otherwise.}
\end{cases}
\end{equation}
If both margins are negative the outcome is \textsf{DENY-STALE}, so
\textsf{DENY-LATE} denotes a late decision whose evidence was otherwise fresh.
Table~\ref{tab:outcomes} gives the outcomes. The contribution is not the
conjunction, which is unsurprising once stated, but that the outcome names the
reason: three operationally distinct failures that a binary gate reports
identically, or in the stale case does not report at all.

\begin{table}[t]
\centering
\caption{Outcomes of the assurance contract, and what produces each on the
platform.}
\label{tab:outcomes}
\setlength{\tabcolsep}{11pt}
\renewcommand{\arraystretch}{1.05}
\begin{tabular}{@{}ll>{\raggedright\arraybackslash}p{4.9cm}@{}}
\toprule
\textbf{Outcome} & \textbf{Violated} & \textbf{Example condition} \\
\midrule
\textsf{PERMIT}         & nothing & normal operation \\
\textsf{DENY-INTEGRITY} & $\mathit{integrity}$ & rogue on-board unit; modified decision-point host \\
\textsf{DENY-STALE}     & $\MF$ & periodic re-attestation, request late in the interval \\
\textsf{DENY-LATE}      & $\MR$ & contention exhausting the budget \\
\bottomrule
\end{tabular}
\end{table}

\subsection{Interaction Between Freshness and Timeliness}
\label{sec:masking}

The margins are measured from different origins, but for \emph{held} evidence
they are not independent, and the dependence constrains how $\Fmax$ may be
chosen. Call evidence held when it was generated before the request, which is the
case under periodic attestation. Evidence age at the decision is $\td - \ta$, and
held evidence satisfies $\ta \le \tr$, so $\td - \ta \ge \td - \tr$. Suppose a
decision is late, meaning $\td - \tr > \Dbud(x_r)$, and suppose the freshness
bound satisfies $\Fmax \le \Dbud(x_r)$. Then
\begin{equation}
\td - \ta \;\ge\; \td - \tr \;>\; \Dbud(x_r) \;\ge\; \Fmax ,
\end{equation}
so the evidence is stale as well. By the precedence of
Eq.~\eqref{eq:outcome} every such decision is then reported as
\textsf{DENY-STALE}, and \textsf{DENY-LATE} is unobservable.

Under periodic attestation \textsf{DENY-LATE} is therefore reachable only when
$\Fmax > \Dbud(x_r)$, and the threshold is an infimum rather than a minimum,
since equality still masks. The freshness bound has to be chosen against the
budget rather than in isolation, and since the budget varies with the physical
state, a single fixed $\Fmax$ makes the outcome vocabulary state-dependent: the
same deployment reports differently at different approach speeds.

The premise is exactly what fails under on-demand refresh. Evidence obtained in
response to a request satisfies $\tr < \ta \le \td$, so its age at the decision
can be small while the decision is late, and \textsf{DENY-LATE} becomes
observable at any $\Fmax$. Masking is a property of periodic held evidence and
not of the contract, which is a further reason an on-demand
refresh policy would change the character of a deployment and not only its
performance. The configuration evaluated in Sect.~\ref{sec:results} uses periodic
evidence with $\Fmax = \Dbud(x_r)$, which is the masking case, and the results
report no \textsf{DENY-LATE} accordingly.

\subsection{Runtime Monitor Realisation}
\label{sec:monitoring}

The obligation is timed and its bound is fixed by data at the triggering event. That is not a metric-temporal formula with constant interval bounds, so it
is not directly expressible in metric first-order temporal
logic~\cite{monpoly2012}, where interval bounds are constants rather than terms
over first-order variables. We therefore compute the absolute deadline in the
instrumentation layer and carry it as data. The requester emits
$\textit{Request}(r, \tr, \tr + \Dbud(x_r))$ and the verifier emits
$\textit{Decision}(r, \td, \ldots)$, so the check becomes first-order arithmetic
over data terms and event order rather than a variable metric interval.

No engine cross-check is reported. Once the deadline travels as data the
obligation reduces to arithmetic over recorded terms and an ordering constraint,
so discharging it on a temporal engine such as MonPoly~\cite{monpoly2012} would
validate the encoding rather than the contract.

\section{HIL Platform and Experimental Method}
\label{sec:platform}

The platform couples a driving simulator to real attestation hardware, so that
physical state is simulated and assurance behaviour is measured. This section
gives the components, the two verification pillars, where every timestamp is
taken, and how contention is generated and reported.

\begin{figure}[tbp]
\centering
\resizebox{\linewidth}{!}{%
\begin{tikzpicture}[font=\scriptsize,>=Latex,node distance=6mm,
  box/.style={draw,rounded corners=1pt,minimum height=9mm,minimum width=2.5cm,align=center},
  hw/.style={box,fill=black!8,very thick},
  sim/.style={box,fill=black!2,dashed}]

\node[sim] (carla) {\textbf{Simulation host}\\CARLA\\vehicle, geometry, signal};
\node[box,right=16mm of carla] (broker) {\textbf{Broker}\\Mosquitto, MQTT QoS 0};
\node[hw,right=16mm of broker] (rsu) {\textbf{Roadside unit}\\Fedora, TPM 2.0\\IMA $+$ verifier};
\node[hw,below=9mm of broker] (obu) {\textbf{On-board unit}\\ESP32\\MAC $+$ firmware hash};
\node[box,below=8mm of rsu,fill=black!4] (load) {\textbf{Load generator}\\graded, calibrated};

\draw[->] (carla) -- node[above]{beacon} (broker);
\draw[->] ($(broker.east)+(0,1.4mm)$) -- node[above]{request}
      ($(rsu.west)+(0,1.4mm)$);
\draw[->] ($(rsu.west)-(0,1.4mm)$) -- node[below]{decision}
      ($(broker.east)-(0,1.4mm)$);
\draw[<->] (obu) -- (broker);
\draw[->,dashed] (load) -- (broker);

\begin{scope}[on background layer]
\node[draw,dotted,thick,inner sep=3.5mm,fit=(obu)(rsu)(load),
      label={[font=\scriptsize\itshape,text=black!60]below:measured, not simulated}] {};
\end{scope}
\node[font=\scriptsize\itshape,text=black!60,below=1.5mm of carla] {simulated};
\end{tikzpicture}}
\caption{Hardware-in-the-loop topology. The dotted region is real: a physical
microcontroller, a TPM-backed host running Linux integrity measurement, and the
contended transport between them. Only the vehicle and the intersection are
simulated.}
\label{fig:topology}
\end{figure}
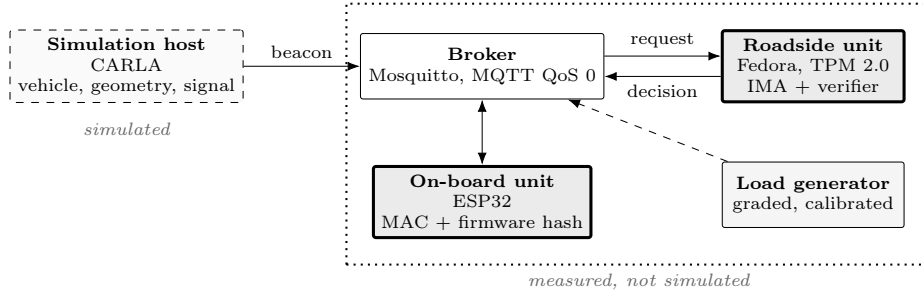

\subsection{Hardware-in-the-Loop Platform}

\begin{figure}[ht]
\centering
\begin{minipage}[c]{0.575\linewidth}\centering
\includegraphics[width=\linewidth]{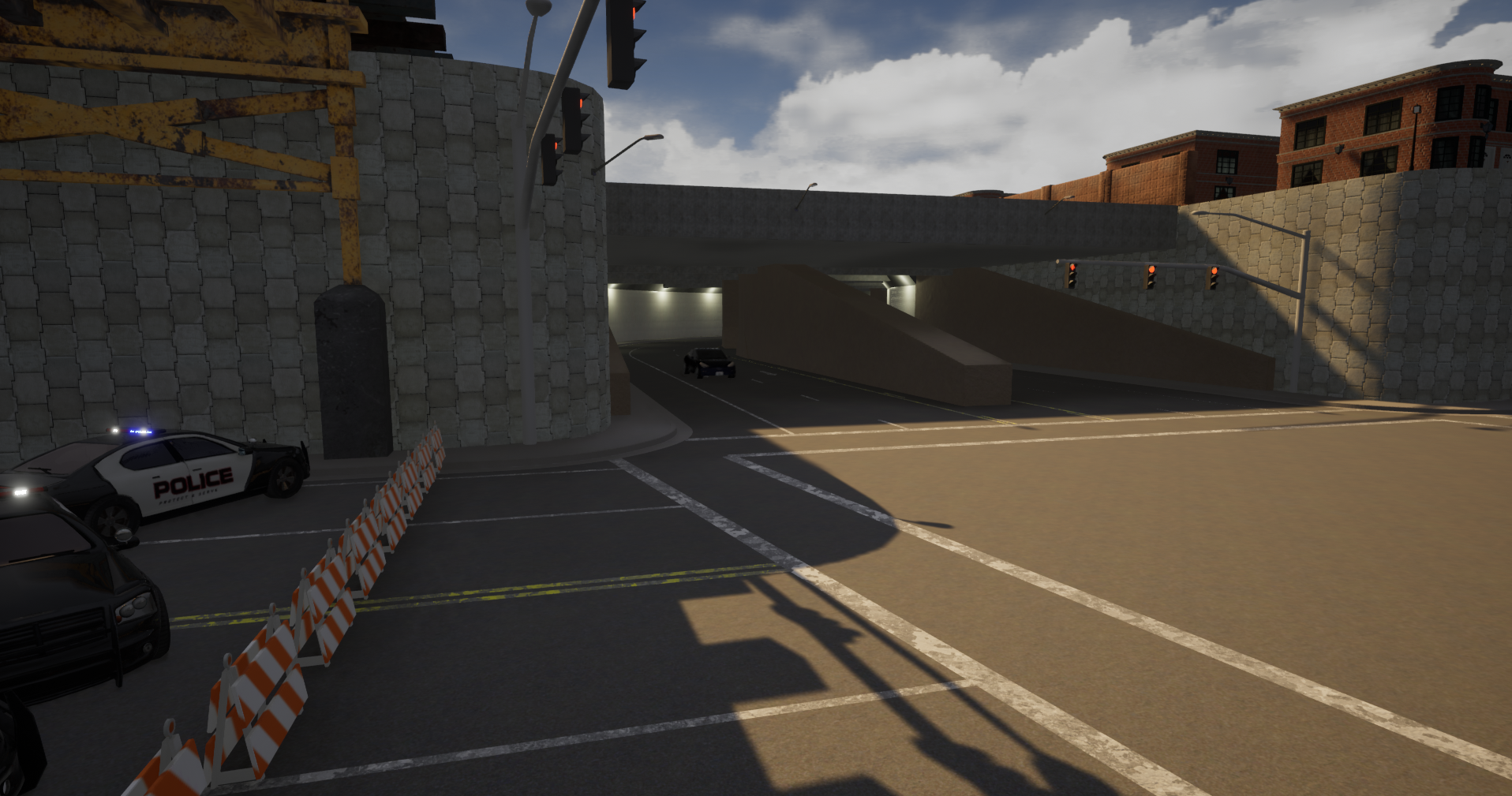}\\[3pt]
{\scriptsize (a) the simulated approach in CARLA}
\end{minipage}\hfill
\begin{minipage}[c]{0.395\linewidth}\centering
\includegraphics[width=\linewidth]{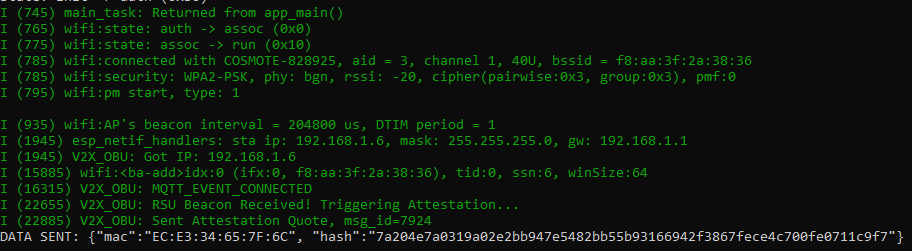}\\[3pt]
{\scriptsize (b) the on-board unit under test}
\end{minipage}
\caption{The physical state is simulated and the assurance evidence is measured.
CARLA supplies distance and speed at the moment of the request; the ESP32
supplies the admission payload that the roadside unit checks.}
\label{fig:rig}
\end{figure}

Figure~\ref{fig:topology} gives the arrangement, and Fig.~\ref{fig:rig} shows the
two halves of it. The division follows the usual hardware-in-the-loop
pattern~\cite{isermann1999}: the plant is simulated and the components under
study are real, here the attestation stack rather than a controller. The simulation host runs CARLA~\cite{carla2017} on
Windows, which spawns the
vehicle, advances it towards the intersection, detects entry into the trigger
volume and actuates the signal on command. The on-board unit is a physical ESP32
microcontroller that listens for the proximity beacon and answers with its
registered hardware address and a SHA-256 digest of its firmware, recomputed at
each request rather than stored. The
roadside unit is a Fedora host with a TPM 2.0, running the Linux Integrity
Measurement Architecture~\cite{ima2004} under a Keylime
verifier~\cite{keylime2016,keylimedsn2025}. A Mosquitto broker carries all traffic using MQTT~\cite{mqtt2019}.

Quality of service is not uniform across the loop: the on-board unit publishes at
level~1 while everything else publishes and subscribes at level~0, so one
acknowledgement round trip sits inside every end-to-end measurement reported
here.

\paragraph{What the simulator supplies.}
The contract needs only distance and speed, which a kinematic model would also
supply, so it is worth saying what the simulator adds beyond two scalars. The
trigger geometry is derived rather than assumed: traffic-light actors sit on the
far side of the junction in these maps, so the reference point is transformed
from the actor location to the world coordinates of the trigger volume, and a
request issued against the actor position fires only after the vehicle has
entered the intersection. Actuation executes against a signal object the
simulator holds, so a \textsf{PERMIT} changes state and a refusal leaves the
intersection safe. And the plant admits several vehicles with distinct identities
on one approach, the route to the concurrent-request experiments
Sect.~\ref{sec:threats} records as outstanding.

\paragraph{Trust topology.}
\label{sec:topology}

Where each part of the attestation stack runs bounds what the second pillar
establishes. The measured agent, the
verifier that evaluates its quotes against the policy, and the decision logic
that consumes the resulting verdict all run on the same Fedora host: the agent
and the verifier are reachable at loopback addresses on that machine, and the
decision logic is a process beside them.

The consequence is a bound on the claim, and we state the narrowed form rather
than the general one. What the second pillar establishes on this platform is the
\emph{detection of modification to files covered by the measurement policy, under
a decision process that is not itself adversarial}. That is what the tamper
experiment of Sect.~\ref{sec:res:discrimination} exercises and it is all it
exercises. The stronger property, that a compromised decision point cannot grant
what it should refuse, is not established here, because a host that has been
taken over can suppress or ignore a verdict about itself before it reaches the
decision. A TPM protects the integrity of a quote; it does not oblige a
compromised consumer to act on one.

Establishing the stronger property requires the consumer of the verdict to sit
outside the attested boundary, the split the Keylime architecture is built for
and which this platform does not instantiate. Either a verifier on a separate
host, so a compromise of the roadside unit cannot reach the process judging it,
or actuation gated at the signal controller on a verdict it obtains
independently, so suppression fails closed. Both are deployment changes and neither touches
the contract; nothing in Sects.~\ref{sec:contract}--\ref{sec:results} depends on
the co-location.

\paragraph{Two verification pillars.}

Edge verification is an admission check: it decides whether the requester is a
registered unit, by comparing the presented hardware address and firmware digest
against a table of admitted pairs. The device has no discrete root of trust, a hardware address is spoofable, and a
digest recomputed by firmware that is itself under an adversary's control can be
replaced by the expected value, so the check establishes admission rather than
cryptographic identity. Spoofing and replay of the on-board credential are therefore outside the threat
model of Sect.~\ref{sec:threat}. This pillar supplies the request-fresh half of
the verification state, against which the periodic half is contrasted. Infrastructure verification answers whether the deciding platform is
itself sound, by binding integrity measurements of the running system to the TPM
and evaluating them against a policy. We run a strict policy covering the trusted
computing base, so a modification to a covered file is caught at the next
measurement rather than at the next reboot.

The second pillar is what makes evidence age a first-class quantity. Edge
verification is performed per request and its result is as fresh as the request.
Infrastructure verification is periodic, its result as fresh as the last cycle.

\subsection{Timing and Evidence-Age Instrumentation}
\label{sec:timing}

Every duration is a difference between two readings of one clock at one
observation point, wall clock in the pilot measurements reported here and
monotonic in the harness that supersedes it. End-to-end latency is stamped on the
simulation host, once immediately before the beacon is published and once on
arrival of the decision. That interval starts one message earlier than $\tr$, so
the measured distribution is an upper bound on $\td - \tr$ and the envelopes
built from it are conservative in that respect. Component costs are stamped on the roadside unit, again
from a single clock. No duration is ever formed by subtracting a timestamp taken
on one machine from a timestamp taken on another, so no result depends on clock
synchronisation between the three hosts and none inherits the residual offset
that even a disciplined protocol leaves~\cite{ieee1588}. Evidence age is the one
exception, since the timestamp it uses is generated by the attestation subsystem
rather than by the observation point; Sect.~\ref{sec:threats} records the
resolution that limits it.

End-to-end latency decomposes into request transport, identity checking, the
status query, policy evaluation and response transport. Two quantities are
instrumented here, the end-to-end duration and the status-query duration; the
remaining terms are not separated, and Sect.~\ref{sec:threats} records the
consequence.

\paragraph{Evidence age.}

The roadside unit records the timestamp of the evidence behind each decision, not
only its verdict, by reading the verifier's record of the last completed
attestation. Where that timestamp is unavailable it is recorded as missing rather
than defaulted to the current time, since defaulting would report an age of zero
and silently destroy the quantity the study is about.

\paragraph{Contention generation.}

Contention is generated on the shared segment by a publisher issuing
$1024$-byte messages at quality of service~0 as fast as the broker will accept
them, and it is reported as measured throughput rather than as a generator
setting. The saturating condition used for the loaded measurements sustained
between $15.9$ and $18.3$ thousand messages per second over $4.4$ million
messages and $259$ seconds, a rate of approximately $17$ thousand messages per
second or $139$ megabits per second at the application layer. Those figures replace a percentage, which describes the generator.

\paragraph{Two verifier interfaces.}

The verifier status was originally obtained by invoking the tenant command-line
utility as a subprocess, which pays process start-up and output parsing on every
request. We replaced it with a direct call to the verifier's REST interface. The
change is engineering rather than research, and its effect on end-to-end cost is
large enough that Sect.~\ref{sec:res:latency} reports it.

\subsection{Experimental Factors and Configurations}
\label{sec:factors}
\label{sec:measures}
\label{sec:method}

The platform measures the quantities the contract is defined over: the timing of
TPM-backed measurement and of the status query, the age of the evidence each
decision consumes, the identity check, the transport under contention, and the
effect of the verifier interface on end-to-end cost. Together they supply the
empirical distribution of $\td - \tr$ and the realisable values of $\Ae$, the two
inputs the envelope is built from.

The plant supplies the other input, distance and speed read at the request, which
makes $\Dbud$ per-request and lets the space be swept at geometries impractical
to instrument on a road.

Trigger distance $20$, $50$, $100$ and $150$\,m; approach speed $50$, $80$ and
$110$\,km/h; evidence-age phase $\phase$ at $0$, $0.25$, $0.5$, $0.75$ and $1$ of
the attestation interval; and attestation interval $0.5$, $1$, $2$, $5$ and
$10$\,s. That last range is not
arbitrary. The verifier configuration reports an attestation period of $2$\,s
and a maximum tolerated interval of $10$\,s, so both endpoints are values the
deployed system permits. Latency measurements additionally vary the verifier
interface between the command-line utility and REST.

Evidence age is sampled as a \emph{phase} of the attestation interval rather than
in absolute seconds, so the result survives a change of interval:
$\phase = (\tr - \ta)/T$.

\subsection{Metrics and Statistical Analysis}
\label{sec:metrics}

Latency is reported as median, p95 and maximum observed alongside the mean. Contract-satisfaction fractions are point estimates over the measured latency
samples. Wilson score intervals are computed for each and retained with the
data; none is reported here, because the criterion ranks operating points rather
than certifying them.
No execution-time or network bound has been established, so every extreme quoted
is a maximum observed value and none is a worst case.

\section{Results}
\label{sec:results}

The four questions of Sect.~\ref{sec:intro} are answered in order.
Sect.~\ref{sec:res:latency} measures the assurance loop and tests contention
(RQ1); Sect.~\ref{sec:res:envelope} builds the envelope (RQ2) and compares it
against the one a security-blind model admits (RQ3);
Sect.~\ref{sec:res:interval} varies the interval and the latency distribution to
see which moves it further (RQ4). A final subsection asks what the contract
distinguishes that a binary gate does not.

\subsection{Assurance Latency and the Effect of Contention}
\label{sec:res:latency}

Table~\ref{tab:latency} reports the measured latency distribution the envelope is
built from. Under saturating channel contention the mean rises to $538.7$\,ms, an
increase of $27.4\%$ on the unloaded mean.

\begin{table}[t]
\centering
\caption{The measured latency distribution, without channel contention, from
which $Q(q)$ is taken. Extremes are maxima observed rather than bounds. The loaded arm survives as summary statistics only, used in
Sect.~\ref{sec:res:latency} for the existence test alone.}
\label{tab:latency}
\setlength{\tabcolsep}{7pt}
\renewcommand{\arraystretch}{1.15}
\small
\begin{tabular}{@{}lrrrrr@{}}
\toprule
\textbf{Configuration} & $n$ & \textbf{mean} & \textbf{median} & \textbf{p95} & \textbf{max obs.} \\
\midrule
End-to-end, no contention   & 50 & 422.8 & 419.8 & 559.2 & 624.2 \\
Attestation component       & 50 &  33.3 &  20.0 &  93.9 & 147.3 \\
\bottomrule
\end{tabular}
\end{table}

The loaded condition was collected over seven trials against fifty unloaded, and
survives only as summary statistics with no retained samples. That is enough for
a test and not enough for a magnitude, and the two questions separate cleanly.

\paragraph{Is there a contention effect at all?}
Under exchangeability, the exact rank test gives the probability that the
smallest of the $n$ loaded observations exceeds $k$ of the $m$ unloaded ones
as $\binom{m+n-k}{n}\big/\binom{m+n}{n}$, which
needs nothing from the loaded arm but its minimum and its count.
Table~\ref{tab:loadedtest} applies it with $m=50$ and $n=7$. End-to-end, the
loaded minimum of $438.67$\,ms exceeds $31$ of the $50$ unloaded samples,
giving
$p = 2.5\times10^{-3}$: a contention effect is supported even at seven trials. On the attestation component the loaded minimum exceeds only one of the fifty,
one further observation being equal to it, and the same test returns
$p = 0.88$.

\begin{table}[t]
\centering
\caption{Exact rank test on the loaded arm: its minimum against the fifty
unloaded samples. Exceedance is strict, so the one observation equal to the
attestation minimum is not counted.}
\label{tab:loadedtest}
\setlength{\tabcolsep}{9pt}
\renewcommand{\arraystretch}{1.05}
\small
\begin{tabular}{@{}lrrrl@{}}
\toprule
\textbf{quantity} & \textbf{loaded min} & \textbf{exceeds} & \textbf{$p$} & \textbf{verdict} \\
\midrule
End-to-end            & 438.67\,ms & 31 / 50 & $2.5\times10^{-3}$ & supported \\
Attestation component & 14.40\,ms  & \phantom{0}1 / 50 & $0.88$ & no evidence \\
\bottomrule
\end{tabular}
\end{table}

The measurements support an end-to-end contention effect
while providing no evidence of a corresponding shift in the attestation
component. Failing to reject is not evidence of zero, and seven observations give
little power against a small effect, so what follows is that the demonstrable
sensitivity lies outside the attestation computation.

\paragraph{How large is it?}
Not answerable here. A seven-sample mean is a fragile point estimate, so the loaded mean
serves only to index one reference row of Table~\ref{tab:sensitivity}. The attestation component is also heavily
right-skewed, with a median of $19.95$\,ms against a mean of $33.30$\,ms and a
maximum observed $147.30$\,ms, a mean-to-median ratio of $1.67$; on a
$10{,}000$-resample bootstrap its unloaded mean has a $95\%$ interval of
$[26.01, 41.70]$\,ms, which contains the loaded mean. Distributional statistics
are reported for the unloaded arm only.
Equal-scale arms under graded contention would give the magnitude, and a per-term
decomposition would locate it precisely. Sect.~\ref{sec:res:interval} tests how
strongly plausible changes in decision latency and in the attestation interval
move the resulting envelope.

\subsection{Constructing the Attestation Envelope}
\label{sec:res:freshness}
\label{sec:res:envelope}
\label{sec:res:gap}

Evidence age enters the contract through the freshness margin. Under periodic
attestation the attainable range of that age is bounded by the attestation
interval, while the age presented to a particular request is set by its phase
within the interval. A vehicle covers $27.8$\,m at $50$\,km/h and $61.1$\,m at $110$\,km/h while
evidence ages through the configured period, and five times those distances
through the tolerated maximum.
The phase is treated analytically rather than measured, for an instrumentation
reason: the verifier exposes the timestamp of the last successful attestation at
one-second granularity, so evidence age cannot be resolved more finely than that
from the verifier alone. Measuring the phase directly needs an agent-side
timestamp, which Sect.~\ref{sec:threats} records as a limitation. With the
configured re-attestation interval, evidence age at the moment of the decision
corresponds to a distance the vehicle has already covered since the platform was
last shown to be trustworthy, and at the top of the interval that distance exceeds
the trigger geometry the deployment was designed around.

The two budgets are of different orders in these units. At the worst phase,
$\phase = 1$, the interface change reduces the decision-latency contribution but
cannot recover the $T$ seconds of evidence age already accumulated before the
request, and that term keeps the required geometry above the deployed $20$\,m at
every speed tested. At $\phase = 0$ the requirement falls to $17.1$\,m at $110$\,km/h.

The envelope below is computed from the fifty unloaded samples and is therefore a
complete result for the no-contention condition rather than a placeholder. What
it does not cover is the contention axis, which Sect.~\ref{sec:res:sensitivity}
addresses by bounding rather than by measurement.

We evaluate the contract over the operating space at $60$ points, four trigger
distances by three speeds by five phases, under two models. The security-blind
model $\Eideal$ assumes assurance is instantaneous and evidence always fresh,
which is what a simulation of vehicle and intersection alone represents. The
hardware-informed model $\Ehil$ substitutes the measured latency distribution and
the evidence age the attestation interval implies. $\Eideal$ admits every point
as feasible
under an empirical $95\%$ satisfaction criterion, meaning that at least that
share of the measured latency samples satisfies the contract at the point. The
criterion is empirical and not a confidence statement: with fifty samples even a
perfect result does not certify an underlying success probability above $95\%$ at
conventional confidence, so it ranks operating points rather than certifying
any of them. $\Ehil$ leaves $76.7\%$ feasible, with losses concentrated where the geometry is
tight: at $110$\,km/h and $20$\,m the contract holds only at phase zero.

That $\Eideal$ admits everything is true by construction rather than by
measurement, so the $23.3\%$ difference restates the infeasible count rather
than adding to it. What it does say is what a simulation of vehicle and
intersection alone omits. The substantive question is how much each omitted
quantity contributes, and Sect.~\ref{sec:res:interval} varies the interval and the
latency separately to answer it.

\paragraph{The boundary in closed form.}
\label{sec:res:boundary}
The grid gives a count, and a count depends on where the grid is sampled. The
boundary itself does not, and under the evaluated configuration it has a closed
form. With $\Fmax = \Dbud$ the freshness test binds, so the contract holds for a
sample exactly when $\phase T + L \le \Dbud$. A point is feasible under the
empirical criterion when at least a fraction $q$ of samples satisfy that, which
is $\Dbud - \phase T \ge Q(q)$ for $Q(q)$ the $q$-quantile of the measured
latency. Substituting $\Dbud = d/v$ gives
\begin{equation}
\label{eq:boundary}
d_{\min}(v) \;=\; v \cdot \bigl(Q(q) + \phase T\bigr),
\end{equation}
a line through the origin in the speed-distance plane whose slope is the total
time the geometry must supply. It reproduces the sixty-point grid classification
exactly.

\begin{table}[t]
\centering
\caption{Minimum trigger distance from Eq.~\eqref{eq:boundary},
$Q(0.95) = 559.2$\,ms.}
\label{tab:boundary}
\setlength{\tabcolsep}{12pt}
\renewcommand{\arraystretch}{1.05}
\begin{tabular}{@{}lrrrr@{}}
\toprule
\textbf{configuration} & \textbf{slope} & \textbf{50\,km/h} & \textbf{80\,km/h} & \textbf{110\,km/h} \\
\midrule
$\phase=0$, $T=2$\,s    & 0.56\,s  &  7.8\,m  &  12.4\,m &  17.1\,m \\
$\phase=0.5$, $T=2$\,s  & 1.56\,s  & 21.7\,m  &  34.6\,m &  47.6\,m \\
$\phase=1$, $T=2$\,s    & 2.56\,s  & 35.5\,m  &  56.9\,m &  78.2\,m \\
$\phase=1$, $T=10$\,s   & 10.56\,s & 146.7\,m & 234.6\,m & 322.6\,m \\
\bottomrule
\end{tabular}
\end{table}

Equation~\eqref{eq:boundary} also makes the structure of the evaluation explicit.
Every envelope number here, the sixty-point counts and both sensitivity sweeps
alike, is a deterministic function of one empirically measured latency
distribution, summarised at the chosen criterion by $Q(q)$, together with
configured parameters. The grid is presentation and the measured distribution is
the evidence, which is the right place for the weight to sit. That distribution
is of end-to-end assurance latency on a platform whose verifier queries a
physical TPM, whose measurements come from a running kernel under a strict
policy, and whose transport traverses a real shared network segment. It is not derivable from a specification, not
obtainable from a simulation of the vehicle and intersection, and not transferable
between deployments. Measuring it is what the hardware is for; the closed form
turns one measurement into an operating region.

Table~\ref{tab:boundary} is the deployment form of the result. The decision cost
contributes a fixed $0.56$\,s and the interval contributes $\phase T$, so at the
configured period a worst-phase request needs a trigger at $78$\,m to be
authorised at $110$\,km/h, against $17$\,m with always-fresh evidence, and
$323$\,m at the tolerated maximum. Trigger geometry becomes a derived quantity
rather than a choice validated afterwards.

\subsection{Sensitivity to Attestation Interval and Decision Latency}
\label{sec:res:interval}

Two parameters can move the envelope: the interval at which evidence is
refreshed, and the latency of the decision that consumes it. This subsection
varies each in turn and then compares them.

\paragraph{Attestation interval.}
The evidence-age bound is a configuration value, and the verifier reports two of
them: an attestation period of $2$\,s and a maximum tolerated interval of
$10$\,s. Both are values the deployed system permits, so both are legitimate
points at which to evaluate the contract. Table~\ref{tab:interval} sweeps between
them.

\begin{table}[t]
\centering
\caption{Feasible operating points against the evidence-age bound, under an empirical $95\%$
satisfaction criterion over the sixty-point space, with $\Fmax = \Dbud(x_r)$,
justified in Sect.~\ref{sec:contract}.
The verifier reports a configured period of $2$\,s and a tolerated maximum of
$10$\,s; the two shorter intervals are included to show the trend below the
configured value. Every infeasible point fails on freshness and none on
timeliness, which is the masking result of Sect.~\ref{sec:masking} rather than
an empirical finding: at $\Fmax = \Dbud$ the freshness test subsumes the
response test, so no point can fail the latter alone.}
\label{tab:interval}
\setlength{\tabcolsep}{12pt}
\renewcommand{\arraystretch}{1.15}
\begin{tabular}{@{}rrrcl@{}}
\toprule
\textbf{interval} & \textbf{feasible} & \textbf{gap} & \textbf{infeas.} & \textbf{note} \\
\midrule
0.5\,s  & 54 / 60 & 10.0\% &  6 & below the configured period \\
1.0\,s  & 52 / 60 & 13.3\% &  8 & below the configured period \\
2.0\,s  & 46 / 60 & 23.3\% & 14 & configured period \\
5.0\,s  & 35 / 60 & 41.7\% & 25 & between the two \\
10.0\,s & 24 / 60 & 60.0\% & 36 & maximum the verifier tolerates \\
\bottomrule
\end{tabular}
\end{table}

At the tolerated maximum the envelope falls to $40\%$ of the space, and
every point lost beyond the nominal period is lost to evidence age with none lost
to latency. In distance terms the vehicle covers $61.1$\,m while evidence ages
through the nominal period at $110$\,km/h, and $305.6$\,m through the tolerated
maximum.

Every point lost beyond the configured period is lost on freshness, which the
comparison below sets against what latency alone can do.

\paragraph{Decision latency.}
\label{sec:res:sensitivity}

The graded contention runs are outstanding, so their effect is bounded rather
than estimated. Scaling the measured unloaded distribution by a factor and
recomputing gives Table~\ref{tab:sensitivity}, which separates points lost
because the evidence was too old from points lost because the decision was too
slow.

\begin{table}[t]
\centering
\caption{Sensitivity of the envelope to a modelled contention slowdown, at the
nominal $2$\,s interval. The scaling is applied to the measured unloaded
distribution; the $1.27$ row is the ratio of the pilot's loaded and unloaded
means, shown for reference only. An infeasible point is \emph{age-only} when it remains
infeasible with latency set to zero, and \emph{age $+$ latency} otherwise. The first column is constant by construction, since that test does not
involve the scaling. No point is infeasible on latency alone; see
Table~\ref{tab:interval}.}
\label{tab:sensitivity}
\setlength{\tabcolsep}{14pt}
\renewcommand{\arraystretch}{1.15}
\begin{tabular}{@{}rrcc@{}}
\toprule
 & & \multicolumn{2}{c}{\textbf{infeasible}} \\
\cmidrule(l){3-4}
\textbf{factor} & \textbf{feasible} & \textbf{age-only} & \textbf{age $+$ latency} \\
\midrule
1.00 & 46 / 60 & 9 &  5 \\
1.27 & 44 / 60 & 9 &  7 \\
1.50 & 43 / 60 & 9 &  8 \\
2.00 & 41 / 60 & 9 & 10 \\
3.00 & 35 / 60 & 9 & 16 \\
5.00 & 27 / 60 & 9 & 24 \\
\bottomrule
\end{tabular}
\end{table}

\paragraph{Where the response margin binds.}
\label{sec:res:mrbind}
One property of the measurements shapes every result above. The largest latency
observed without contention is $624.2$\,ms and the smallest authorisation budget
on the grid is $654.5$\,ms, at a $20$\,m trigger and $110$\,km/h, so no sampled
decision is late at any operating point and $\MR \ge 0$ throughout. Every
infeasible point fails on freshness, and does so because the data contain no late
decisions rather than only because Sect.~\ref{sec:masking} would mask them. The
response margin is therefore inert at these geometries, and the envelope results
are freshness results. The boundary is close: the largest latency observed under
saturating contention, $656.2$\,ms, exceeds that same budget, so the tightest
geometry sits close to the observed boundary at which lateness begins. Extending the grid below $20$\,m would exercise the response margin.

\medskip\noindent
Nine operating points are infeasible from held evidence age alone, at every
scaling, because there $\phase T$ already exceeds the budget before any latency
is added. Increasing latency enlarges the set whose freshness failure arises from
the combined $\phase T + L$ term, from five points at the measured distribution
to twenty-four at a fivefold slowdown. Contention therefore matters, and it matters
by compounding with evidence age rather than by exhausting the budget on its own.
A measured contention arm would sharpen where the crossover falls.

\paragraph{Comparison.}
Moving the interval from $0.5$ to $10$\,s costs $30$ feasible points and a
fivefold scaling of the latency distribution costs $19$, a ratio of $1.6$;
restricted to the configured range of $2$ to $10$\,s the interval costs $22$
points and the ratio is $1.2$. The two are of comparable weight in magnitude and
differ in kind. The latency distribution is a property of the hardware and the
transport, and shifting it fivefold is a large intervention. The interval is a
configuration value, and the range swept here is one the deployed verifier
already permits.

\subsection{Failure-Mode Discrimination}
\label{sec:res:discrimination}

Table~\ref{tab:discrimination} is the argument for the contract. A binary gate
returns two values and therefore cannot separate a platform that has been tampered
with from one whose evidence is merely old, and it does not detect the latter at
all.

\begin{table}[t]
\centering
\caption{Binary gate against the assurance contract. The integrity rows are
executed on hardware. The stale row is not a separate experiment: it is the
region Table~\ref{tab:interval} shows the contract refuses and the gate admits.
\textsf{DENY-LATE} does not appear, for the reason given in
Sect.~\ref{sec:masking}.}
\label{tab:discrimination}
\setlength{\tabcolsep}{10pt}
\renewcommand{\arraystretch}{1.05}
\begin{tabular}{@{}llll@{}}
\toprule
\textbf{Scenario} & \textbf{Binary gate} & \textbf{Contract} & \textbf{Source} \\
\midrule
Genuine requester, clean host & permit & \textsf{PERMIT} & executed \\
Rogue on-board unit           & deny   & \textsf{DENY-INTEGRITY} & executed \\
Modified decision-point host  & deny   & \textsf{DENY-INTEGRITY} & executed \\
Valid but aged evidence       & permit & \textsf{DENY-STALE} & Table~\ref{tab:interval} \\
\bottomrule
\end{tabular}
\end{table}

Both integrity rows were executed on the platform rather than modelled. A rogue
unit presenting a hardware address and firmware digest outside the admitted table
is refused on identity, before the verifier is consulted at all, so the
infrastructure pillar is never reached:

\noindent\begin{minipage}{\linewidth}
\begin{lstlisting}
[RSU ATTESTATION ENGINE] Request from OBU: EC:E3:34:65:7F:6C
 -> OBU VERDICT: FAILED. ESP32 MAC/Hash invalid or tampered!
 -> FINAL ACTION: Sending RED light command. Access Denied.
\end{lstlisting}
\end{minipage}

A modification to a file covered by the strict policy on the roadside unit is
caught at the next measurement, before the command it would have influenced could
execute. The verifier names the integrity event rather than reporting a generic
failure, which is what allows the outcome to be attributed:

\noindent\begin{minipage}{\linewidth}
\begin{lstlisting}
"severity_level": 6,
"last_event_id": "ima.validation.ima-ng.not_in_allowlist",
"attestation_count": 0,
"attestation_status": "FAIL"
\end{lstlisting}
\end{minipage}

The rows differ in what they establish. The two integrity rows show the
mechanisms detecting what they are built to detect, a digest mismatch and a
policy-covered file change, under the narrowed claim of
Sect.~\ref{sec:topology}; neither capability is new, and what the contract adds
is attributing both to one outcome while separating them from the third row. The
stale row is derived from the configured interval rather than executed, and no
\textsf{DENY-LATE} row appears. The vocabulary distinguishes four conditions, of
which this study exercises two on hardware, derives one, and leaves one
unreachable at these geometries.

The requester in that second case is genuine and passes edge verification. The
refusal comes from the platform that would have granted the request, which is the
case a single-pillar design cannot express.
The outcome column for those rows records what the contract assigns to the
observed evidence, evaluated offline over the recorded verdicts; the refusals
themselves were produced by the original gate. \textsf{DENY-LATE} is absent because the
platform runs at a freshness bound equal to the budget, under which
Sect.~\ref{sec:masking} shows lateness cannot be observed.

\section{Discussion and Threats to Validity}
\label{sec:discussion}

\paragraph{A security control that can deny the service it protects.}
Under a fail-secure policy a refusal is safe and it is also a denial, and the
results bear on how reachable that denial is. Contention does slow the loop,
detectably so by the test of Sect.~\ref{sec:res:latency}, though the measurements
do not fix by how much, and an adversary who can forge neither identity nor
platform integrity would defeat neither pillar in attempting it.

Sect.~\ref{sec:res:interval} places the two parameters on a comparable
quantitative footing: a fivefold latency scaling removes nineteen feasible
points, while moving the attestation interval from two to ten seconds removes
twenty-two. The comparison quantifies the sensitivity of the envelope to each
parameter and implies nothing about attacker effort or realisability. We did not
attempt an attack and do not claim one; the measurements give the shape of the surface an attack would have to move.

\subsection{Threats to Validity}
\label{sec:threats}

\paragraph{What the integrity pillars establish.}
The narrowed claim of Sect.~\ref{sec:topology} carries into the results: the
tamper row of Table~\ref{tab:discrimination} shows detection of a policy-covered
modification under a non-adversarial decision process, not that a compromised
decision point can be constrained.

\paragraph{One instantiation.}
One intersection, one on-board unit, one verifier, one vehicle at a time. We
claim a methodology demonstrated on a single concrete instantiation. The
numerical boundaries belong to this platform; what generalises is the
formulation, a deadline derived from mission state together with a measured
assurance distribution, and the finding that evidence age and response latency
must be evaluated against that deadline jointly.

\paragraph{Simulator fidelity.}
CARLA supplies the physical state and is not synchronised with or calibrated
against a physical intersection, and it is a simulator rather than a digital
twin. Vehicle dynamics are a constant-speed approach.
Sect.~\ref{sec:measures} states what the platform supplies in its place.

\paragraph{Transport standing in for a radio.}
MQTT over IP carries the traffic, so the contention generated is transport
contention on a shared segment rather than contention for a V2X radio channel.
The direction of the effect should carry over; its magnitude should not be read
across. The quality-of-service asymmetry described in Sect.~\ref{sec:platform}
compounds this: the acknowledgement round trip on the on-board unit's uplink is a
property of this transport and would have no counterpart in a broadcast radio
layer, so the contention sensitivity we localise there is partly an artefact of
the stand-in.

\paragraph{The deadline is an upper bound.}
As discussed above; every envelope reported is optimistic in consequence.

\paragraph{Sample sizes, and what the loaded arm can carry.}
The measurements are a pilot: fifty trials unloaded with samples retained, and
seven loaded with only summary statistics retained. The loaded arm supports the
existence of a contention effect, by the exact test of
Sect.~\ref{sec:res:latency}, and it does not support a magnitude. We use it for
the first and not the second, and no distributional statistic is computed for
it. Larger equal-scale arms would give the magnitude;
Sect.~\ref{sec:res:interval} bounds how much that could move the conclusion.

\paragraph{An unresolved discrepancy in the pilot record.}
The unloaded file carries a summary line reporting an end-to-end mean of
$421.77$\,ms and an attestation mean of $29.17$\,ms. Recomputing from the fifty
rows above that line gives $422.83$ and $33.30$, while minima and maxima agree
exactly. We report the recomputed values throughout, since the samples are the
primary record, and note that the summary was evidently generated from a
different or earlier set.

\paragraph{Evidence age is analytic, not measured.}
The verifier exposes the timestamp of the last successful attestation at
one-second granularity, so evidence age cannot be resolved more finely than that
from the verifier alone, and the phase axis of the envelope is therefore derived
from the configured interval rather than measured per request. Resolving it
directly requires an agent-side timestamp. The endpoints of the interval range
are configured values of the deployed verifier, so the range itself is not an
assumption.

\paragraph{What is not evaluated here.}
Four things. We do not demonstrate
a denial attack, only the contention effect that would underlie one and the
sensitivity analysis that bounds its reach. We do not evaluate the adaptive
refresh policy, which is stated as a consequence of the contract and left to
future work. And we do not decompose latency per term, so the localisation of the
contention effect rests on the one-sided rank test of Sect.~\ref{sec:res:latency}
rather than on a breakdown. Concurrency is likewise not instantiated: the
platform serves one requester at a time.

\section{Related Work}
\label{sec:related}

Five strands bear on this work: attestation architectures and the treatment of
freshness within them, attestation of constrained devices, timeliness as a
security requirement, availability of vehicular communication, and runtime
verification for cyber-physical systems.

\paragraph{Attestation architectures and evidence freshness.}
Freshness has been one of the foundational principles of remote attestation since
the area was first set out systematically~\cite{coker2011}, and the RATS
architecture is explicit that the freshness of both evidence and attestation
results is a matter of policy, discussing expiry and the mechanisms that
establish recency~\cite{rfc9334}. The trust anchor on
general-purpose hosts is a TPM~\cite{tcg2019}, and the measurement discipline
that produces the evidence is IMA~\cite{ima2004}, with continuous verification
against a policy supplied by Keylime~\cite{keylime2016}, whose behaviour in
practice, including the failure modes of continuous attestation, has itself been
studied empirically~\cite{keylimedsn2025}. In this literature
freshness is a first-class concept and a configured quantity. What it is not is a
quantity evaluated against anything outside the attestation system: the recency
requirement is stated in seconds and justified by convention rather than derived
from the process the verdict will be used to control. Our contribution is the
derivation.

\paragraph{Attestation of constrained devices.}
The on-board unit here has no discrete root of trust, which is the common case at
the edge and has produced a substantial literature on software and hybrid
alternatives. Collective schemes attest large device populations at
scale~\cite{seda2015}, and control-flow attestation extends the evidence from
what was loaded to how it executed~\cite{cflat2016}. That work concerns what evidence a constrained
device can produce and at what cost, and is orthogonal to the question here,
which arises after the evidence exists.

\paragraph{Timeliness as a security requirement.}
Closest to this work, ARI formulates real-time mission execution integrity and
argues that autonomous cyber-physical systems require attestation that is timely
as well as correct~\cite{ari2023}. The difference is the direction of the
question and the unit of analysis. ARI attests that an execution proceeded
correctly and within its schedule, so the attested object is the execution; we
ask whether an attestation verdict is still decision-valid for an authorisation
whose admissible response time is derived at run time from physical state, so the
attested object is the platform and the timed object is the decision that
consumes it. More broadly, the tension between security mechanisms and timing
guarantees is a mature line in real-time systems, including frameworks that
schedule security tasks against real-time
guarantees~\cite{contego2017}. Those results assume execution-time bounds and schedulability analysis, neither
of which a TPM-backed verifier on commodity hardware supplies, so we report
measured distributions and never a certified bound.

\paragraph{Vehicular communication, availability and preemption.}
The attack surface automation introduces in vehicles is well
catalogued~\cite{petit2015} and the connected-vehicle stack
surveyed~\cite{coppola2016}. Within signal preemption, communication delay has
been modelled as a first-class input to performance~\cite{evpdelay2018}, and the
transition requirements that bound what a controller may do on receiving a
preemption request are documented in traffic-control
guidance~\cite{fhwa2008}. Our flooding is a stressor drawn from that
setting rather than a contribution, and that literature is why our deadline is an
authorisation-decision budget and not a preemption one.

\paragraph{Runtime verification for cyber-physical and edge systems.}
Monitoring an execution against a temporal specification is the framing we
borrow~\cite{bauer2011}, with the tools
surveyed in depth~\cite{falcone2021} and the difficulties of carrying it beyond
software into cyber-physical settings set out by
S\'anchez et al.~\cite{sanchez2019}. Monitoring metric first-order temporal
properties supplies the engine tradition our obligation would sit in were its
bound constant~\cite{monpoly2012}, and the reasons it is not are given in
Sect.~\ref{sec:monitoring}. Distributing monitors over an edge topology asks where a monitor runs and what it
can observe from there~\cite{tsigkanos2022}, and applying runtime verification to
co-simulated models~\cite{temperekidis2022fmirv} and digital-twin
architectures~\cite{temperekidis2023dt} asks what a monitor sees when part of the
system under observation is a model. Hierarchical deployments combine engines with
complementary strengths across edge tiers~\cite{kekatos2026csr,kekatos2027ares},
make the resilience of the monitoring transport part of the
specification~\cite{kekatos2026fabric}, and extend the discipline to energy
systems~\cite{kekatos2026resilagent} and to embedded vehicular
platforms~\cite{chintri2026edge}. All of that concerns what a monitor
can express and where it can run. The obligation here is small by comparison, and
the envelope it induces once the deadline becomes a function of physical state is
what carries the contribution.

\section{Conclusion}
\label{sec:conclusion}

Integrity evidence is normally treated as a predicate, and it is not one. It is
generated at a moment, it ages, the decision that consumes it costs time, and the
physical process that waits for that decision has a horizon. We formulated
mission-aware attestation as a contract over those three quantities, showed that
the resulting four outcomes separate failures a binary gate reports identically
or not at all, and showed that the two margins interact: below a threshold on the
freshness bound, lateness is unobservable because every late decision is also
stale.

On a hardware-in-the-loop platform in which a simulator supplies the physical
state and a TPM-backed roadside unit supplies the evidence, combining the two
yields a region rather than a number: the part of the operating space in which a
security verdict is still actionable. A security-blind model that omits assurance
behaviour admits a larger envelope than a hardware-informed one does, the
difference is concentrated in evidence age rather than in decision latency.
Sensitivity analysis shows both parameters move the envelope materially: a
fivefold latency scaling removes nineteen feasible operating points, against
twenty-two when the attestation interval moves from its configured two seconds to
the ten the verifier tolerates.

For a time-critical autonomous action, trustworthy evidence is actionable only
while both its age and the time required to consume it remain inside a
mission-dependent attestation envelope. Neither an attestation interval
nor a freshness bound can be chosen from security requirements alone.

\bibliographystyle{splncs04}
\bibliography{references}

\end{document}